\documentclass[twocolumn,prb,showpacs]{revtex4}
\usepackage{amssymb,amsmath}
\pdfoutput=1
\usepackage{graphicx,float}
\usepackage{bm}
\usepackage{array}
\usepackage{algorithm}
\usepackage{algorithmic}

\begin{document}


\title{\bf Spectral properties of hyperbolic nano-networks with
  tunable aggregation of simplexes}
\author{Marija Mitrovi\'c Dankulov$^{2,1}$ ,  Bosiljka
  Tadi\'c$^{1,3}$, Roderick Melnik$^{4,5}$}
\vspace*{3mm}
\affiliation{$^1$Department of Theoretical Physics, Jo\v zef Stefan Institute,
Jamova 39, Ljubljana, Slovenia}
\affiliation{$^2$Scientific Computing Laboratory, Center for the Study of Complex Systems, Institute of Physics Belgrade, University of Belgrade, Pregrevica 118, 11080 Belgrade, Serbia}
\affiliation{$^3$Complexity Science Hub Vienna, Josephstadterstrasse
    39, Vienna, Austria}
\affiliation{$^4$MS2Discovery Interdisciplinary Research Institute; M2NeT Laboratory and Department of Mathematics,
Wilfrid Laurier University, 75 University Ave W, Waterloo, ON, Canada N2L 3C5 }
\affiliation{$^5$BCAM - Basque Center for Applied Mathematics; Alameda de Mazarredo 14, E-48009 Bilbao, Spain}\vspace*{3mm}

\begin{abstract}
\noindent 
Cooperative self-assembly can result in complex nano-networks with new
hyperbolic geometry. However, the relation between the hyperbolicity
and spectral and dynamical features of these structures remains
unclear. Using the model of aggregation of simplexes introduced in I [Sci. Rep., 8:1987, 2018], here we study topological and spectral properties of a large class of self-assembled structures consisting of monodisperse building blocks (cliques of size $n=3,4,5,6$) which self-assemble via sharing the geometrical shapes of a lower order.  The size of the shared sub-structure is tunned by varying the chemical affinity $\nu$ such that for significant positive $\nu$ sharing the largest face is the most probable, while for $\nu < 0$, attaching via a single node dominates.  
Our results reveal that, while the parameter of hyperbolicity remains
$\delta_{max}=1$ across the assemblies, their structure and spectral
dimension $d_s$ vary with the size of cliques $n$ and the affinity
when $\nu \geq 0$. In this range, we findthat $d_s >4$ can be reached
for $n\geq 5$ and sufficiently large $\nu$. For the aggregates of
triangles and tetrahedra, the spectral dimension remains in the range
$d_s\in [2,4)$, as well as for the higher cliques at vanishing  affinity. On the other end, for
$\nu < 0$, we find $d_s\eqsim 1.57$ independently on $n$.
Moreover, the spectral distribution of the normalised Laplacian eigenvalues  has a
characteristic shape with peaks and a pronounced minimum, representing
the hierarchical architecture of the simplicial complexes.
These findings show how the structures  compatible with 
complex dynamical properties can be assembled by controlling the higher-order connectivity among the building blocks.
\end{abstract}

\maketitle

\section{Introduction\label{sec-intro}}
Impact of the system's architecture  onto emergent functionality in
nanostructured materials  has been evidenced by experimental
investigation of the functional features by varying the building
blocks in different self-assembly processes
\cite{SA_colloids_scales2016,SA_collectiveMagnetic_AdvMat2016,Jap2_icosahedrons,AT_Materials_Jap2016,SA_forcesSmall2015,SA_Flowers,SA_hierarchicalMagnPlasmonic_ACSN2011,SA_collectiveNR_SoftMatt2013,Geometry_colloidsPRE2015}.  
Theoretically,  the structural elements that lead to an improved function can be identified by  parallel investigations of
the topology and dynamics of a particular system represented by
nano-network \cite{we_nanonetworks2013}, for example,  by conducting nanoparticle films
\cite{we_NL2007,we_topReview2010,we_SFloops2006,we_Qnets2016}, carbon
nanotube fillers \cite{nano-fillers}, etc.
On a more global scale, the interplay between the structure and dynamics is
captured by spectral properties of networks \cite{Spectra-topologyPRE2017,Spectra-reviewChaos2018}.
More specifically,  spectral analysis of the adjacency matrix or  the
Laplacian operator related to the  adjacency matrix
\cite{Spectra-Lap-Adj} revealed Fiedler spectral partitioning of the
graph and detection of functional modules or mesoscopic communities  \cite{Spectra-wePRE2009,Spectra-commEleonora2018},
hierarchical organisation and homeostatic response
\cite{Spectra-hierarchPRE2005}, the structural changes at the
percolation threshold \cite{Spectra-EVpercoltransNJP2006}, or the occurrence of assortative correlations between nodes \cite{Spectra-assortPRE2015}.
A direct relation between the Laplacian eigenspectrum and the
diffusion processes on that network revealed the role of the
small-degree nodes and features of the return-time of random walks 
\cite{Spectra-SFnetsSergey2003,Spectra-wePRE2009}, as well as the universality of dynamical phase
transitions \cite{Spectra-phtransPRE2018} and 
a deeper understanding of synchronisation on complex networks \cite{Spectra-synchroAlbert2008}.
In this context, the key quantity that relates the structure to the diffusion and synchronisation on a network is the \textit{spectral dimension}
\cite{durhuus2007spectral,Spectra-FTds_Seroussi2018,millan2019synchronization},
which can be determined from the properties of the Laplacian spectrum.

The complex functional systems often exhibit a hierarchical architecture and the related hyperbolic geometry. They can be parameterised by
\textit{simplexes}, cliques of different orders, and described by mathematical techniques of the algebraic
topology of graphs \cite{ATKIN1972,Qanalysis2,cliquecomplexes,jj-book,SM-book2017}.
In the materials science, such structures are grown by cooperative
self-assembly 
\cite{SA_colloids_scales2016,SA_forcesSmall2015,SA_Flowers,SA_chains,SA_hierarchicalMagnPlasmonic_ACSN2011,SA_collectiveNR_SoftMatt2013,Geometry_colloidsPRE2015}. Moreover,
the idea of hierarchical architecture is a center-piece in the
development of many modern innovative technologies such as 3D printing \cite{adv-mat-rev2019}.
Recently, 
these processes have been modelled by attachments of pre-formatted objects
under geometric constraints and specified binding rules
\cite{AT_Materials_Jap2016,we-SciRep2018}. See other similar works in \cite{Geometry_BianconiSR2017,Spectra-rapisarda2019}. 
Whereas, in real complex systems whose structure is detectable from experimental data, the corresponding networks can be decomposed into simplicial complexes. For example,  in brain networks
\cite{we-Brain-PLOS2016,we-Brain-Connectome2019} these simplicial complexes comprise the inner structure of brain anatomical modules.  The hierarchical organisation was also found 
in social networking dynamics \cite{we-PhysA2015,we_Tags2016,Geometry_weEJP2018}, in problems
related to traffic dynamics \cite{we_PRE2015}, and so on.

As mentioned above, the hierarchically organised networks
possess emergent hyperbolicity or negative curvature in the
shortest-path metric, that is they are Gromov hyperbolic graphs
\cite{HB-spaces-decomp2004,HB-BermudoHBviasmallerGraph2013,HB-Bermudo2016,Hyperbolicity_chordality2017,Hyperbolicity_cliqueDecomposition2017}. 
Recently,
the graphs with a small hyperbolicity parameter $\delta$ have been in
the focus of the scientific community for their ubiquity  in real systems and applications, as well as due to their 
mathematically interesting structure
\cite{HB-BermudoHBviasmallerGraph2013,HB-distortionDelta2014,HB-Bermudo2016,Hyperbolicity_chordality2017}. Namely, the upper bound of a small hyperbolicity parameter can be determined from a subjacent smaller graph of a given structure.  
Generally, it is assumed that both naturally evolving, biological,
physical and social systems develop a negative curvature to optimise their  dynamics \cite{Hyperbolicity_IEEE2016,Geometry_BianconiSR2017,salnikov-SCarxiv2018,Geometry_weEJP2018,we-Brain-Connectome2019}.
However, the precise relationship between the hyperbolicity of a network and its spectral and dynamical features remains mostly unexplored.

In this paper, we tackle these issues by systematically analysing the spectral properties of a class of Gromov 1-hyperbolic networks but with the different architecture of simplicial complexes. Based on the model for the cooperative self-assembly of simplexes introduced in \cite{we-SciRep2018}, here we grow several classes of nano-networks and analyse their topology and spectral properties; the monodisperse building blocks are cliques of the order $n=3,4,5,6$ while the geometrical compatibility tuns their assembly in the interplay with the varying  chemical affinity $\nu$ of the growing structure towards the binding group. Specifically, for the negative values of the parameter $\nu$, the effective repelling interaction between the simplexes occurs, while it is gradually attractive for the positive $\nu$. At $\nu=0$ purely geometrical factors play a role. Our results show that while the hyperbolicity parameter remains constant $\delta=1$ for all classes, their spectral dimension varies with the chemical affinity $\nu$ and the size of the elementary building
blocks $n$.  Moreover, these networks exhibit a community structure
when the parameter $\nu\geq 0$.  The inner structure of these
communities consists of simplicial complexes with a hierarchical
architecture, which manifests itself in the characteristic spectral properties of the Laplacian of the network.
 
In Sec.\ \ref{sec-clnets}, we present details of the model and
parameters, while in Sec.\ \ref{sec-topol} we study different topology features of the considered networks. In Sec.\
\ref{sec-spectra} we analyse in detail spectral properties of all
classes of these networks for varied parameters $\nu$ and the size of elementary blocks. Sec.\ \ref{sec-discussion} is devoted to the discussion of the results.

\section{Self-assembly of simplexes and the type of emergent
  structures\label{sec-clnets}}
\begin{figure}[!htb]
\begin{tabular}{cc} 
\resizebox{16pc}{!}{\includegraphics{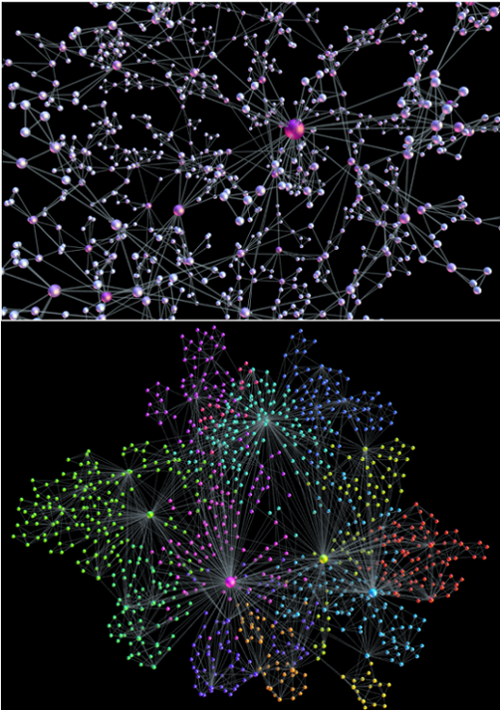}}\\
\end{tabular}
\caption{Aggregates of tetrahedra with strong repulsion, a segment is
  shown in the top panel,  and the case with strong attraction resulting
  in the network with communities is shown in the bottom panel.}
\label{fig-nets-tetra2x} 
\end{figure}

\begin{figure}[!htb]
\begin{tabular}{cc} 
\resizebox{16pc}{!}{\includegraphics{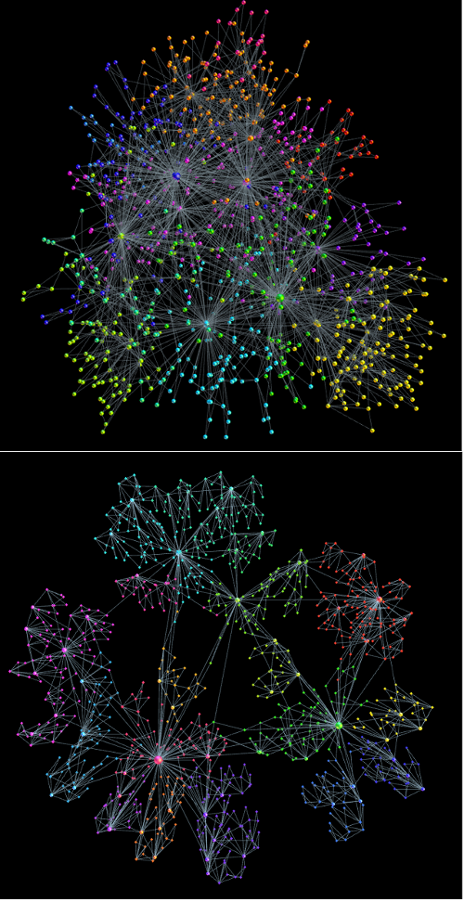}}\\
\end{tabular}
\caption{The networks of the aggregated cliques of mixed 
  sizes $n\in [3,6]$ distributed according to $\varpropto n^{-2}$ for $\nu=
  5$ (top),  and  aggregates of triangles for   $\nu=9$ (bottom). The community
  structure is indicated by different colours of nodes.}
\label{fig-nets-examples} 
\end{figure}

To grow different nano-networks by chemical aggregation of simplexes, we apply the rules of the model for cooperative self-assembly \cite{we-SciRep2018,we_SAloops}. Pre-formatted groups of particles are described by simplexes (full graphs, cliques) of different size $n\equiv q_{max}+1$, where $q_{max}$ indicates the order of the clique. Starting from an initial simplex, at each step, a new simplex is added and attached to the growing network by \textit{docking} along one of its faces, which are recognised as simplexes of the lower order $q=0,1,2,\cdots q_{max}-1$, see online demo \cite{we-ClNets-applet}. 
For example, a tetrahedron can be attached  by sharing a single nod, i.e., a simplex of the order $q=0$ with the
existing network, or sharing an edge, $q=1$, or a triangle, $q=2$,
with an already existing simplex in the network.
The attaching probability depends both on the geometrical
compatibility of the $q$-face of the adding simplex with the current structure as well as on the parameter $\nu$ that describes the chemical affinity of that structure towards the addition of  $n_a=q_{max}-q$ new
vertices.  More precisely, we have \cite{we-SciRep2018}
\begin{equation}
p(q_{max},q;t)= \frac{c_q(t)e^{-\nu (q_{max}-q)}}{\sum _{q=0}^{q_{max}-1}c_q(t)e^{-\nu (q_{max}-q)}} \ 
\label{eq-pattach}
\end{equation}
for the normalised probability that a clique of the order $q_{max}$
attaches along its face of the order $q$. Here,  $c_q(t)$ is the
number of the geometrically similar docking sites of the order $q$ at
the evolution time $t$. Eventually, one of them is selected
randomly. By varying the parameter $\nu$ from large negative to large
positive values, the probability of docking along with a particular
face is considerably changed. For example, for the negative values of
$\nu$, the growing system 'likes' new vertices; consequently, a
simplex preferably attaches along a shared vertex rather than a larger
structure. Effectively, a repulsion between simplexes occurs, see
Fig.\ \ref{fig-nets-tetra2x} top.  In the other limit, for a large
positive $\nu$, the most probable docking is along the potentially
largest face, such that an added simplex of the size $n$ shares the
maximum number $n-1$ of vertices with the existing structure, see
bottom panel in Fig.\
\ref{fig-nets-tetra2x} and  Fig.\
\ref{fig-nets-examples}.   Here, the simplexes in question experience a strong
attraction, which gradually decreases with decreasing $\nu$. For the
neutral case $\nu =0$, the assembly is regulated by strictly geometrical
compatibility factors $C_q(t)$, which change over time as the network grows.

In the original model \cite{we-SciRep2018}, the size of the incoming
simplexes is taken from a distribution, whose parameters can be
varied. To reveal the impact of the size of these building blocks on
the spectral properties of the new structure, here we focus on the
networks with  \textit{monodisperse} cliques, in particular, we
investigate separately the structures grown by aggregation of cliques of the size $n=$3,4,5 and 6 for varied affinity $\nu$. For comparison, we also consider the case with a distribution of
simplexes in the range $n\in[3,6]$.   As the examples in Fig.\
\ref{fig-nets-tetra2x} and Fig.\ \ref{fig-nets-examples} show, the structure of the assembly varies considerably both with the size of simplexes and the level of attraction between them.
Notice that in the case $n=2$ the simplex consists of two vertices
with an edge between them resulting in a simple random tree
graph. Here,  $q_{max}=1$ and all docking faces are single-vertex
sites ($q=0$). Therefore, the probability $p(1,0;t)=1$  is independent
of the value of the parameter $\nu$.
In this work, we consider networks of different number of vertices
$N=$ 1000, 5000, and 10000.

\section{Topological properties of the assembled nano-networks\label{sec-topol}}
\begin{figure}[!htb]
\begin{tabular}{cc} 
\resizebox{16pc}{!}{\includegraphics{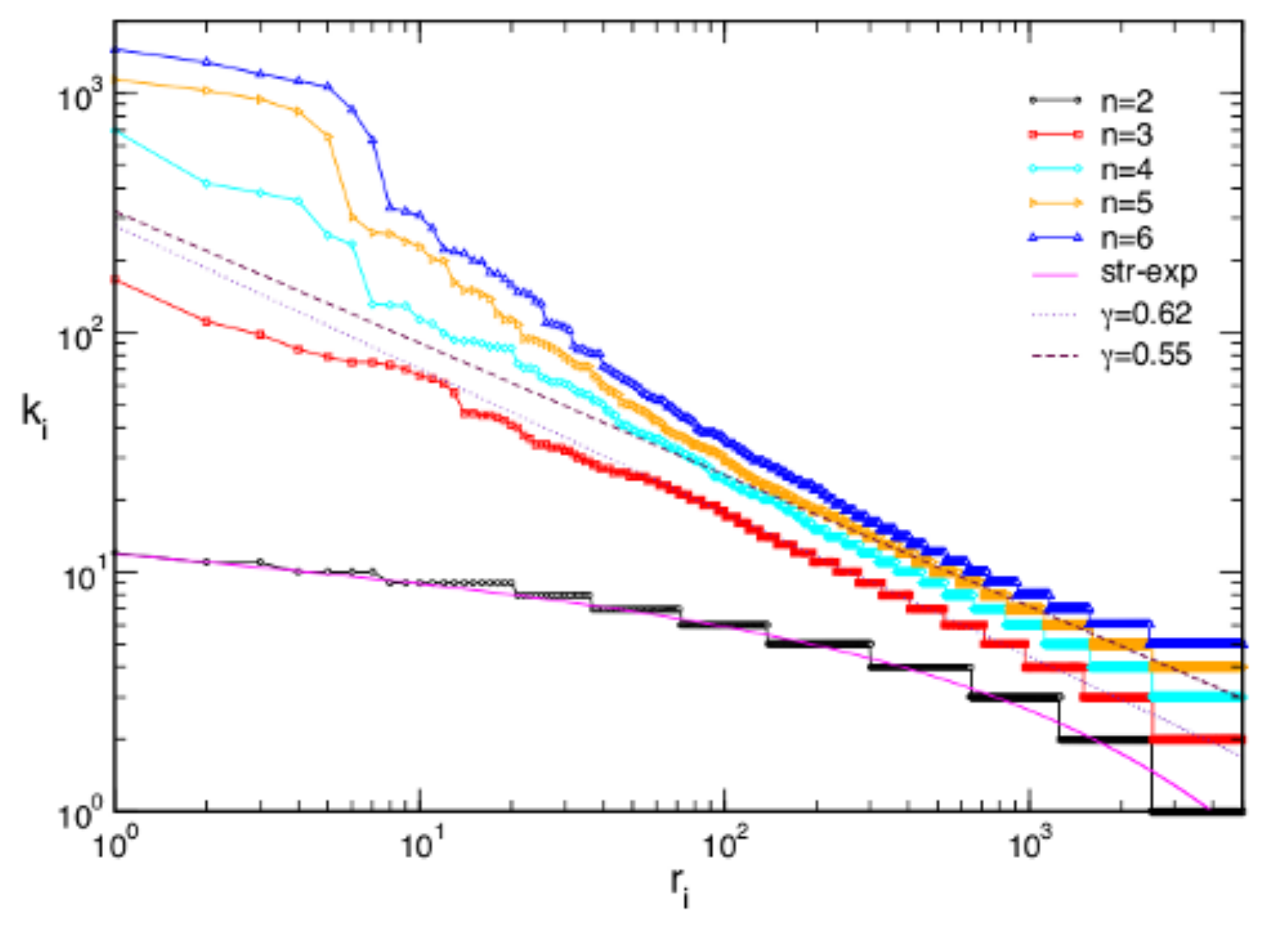}}\\
\end{tabular}
\caption{The degree $k_i$ of the vertex $i$ plotted
  against the vertex rank $r_i$ for different assemblies of cliques of
  size $n$, indicated in the legend,  and $\nu=9$. Stretched
  exponential curve approximates the data for the random tree ($n=2$),
while the asymptotic power-law decay with the exponent $\gamma$ is
appropriate for $n\geq 3$.}
\label{fig-ranking-p9} 
\end{figure}

 The structure of the assemblies strongly depends on the chemical affinity $\nu$ and the size $n$ of the building blocks. For example, a strong repulsion between cliques enables sharing a single node, thus minimising the geometrical compatibility factor and resulting in a sparse graph (a tree-of-cliques). An example with
the tetrahedra as building blocks at $\nu=-9$ is shown in the top panel of Fig.\ \ref{fig-nets-tetra2x}. 
However, for extremally attractive cliques, e.g., for $\nu=9$, the
same building blocks attach mostly via sharing their largest subgraphs
(triangles); thus the geometrical constraints play an important
role. This situation results in a dense network with a nontrivial
community structure, as shown in the bottom panel of Fig.\
\ref{fig-nets-tetra2x} and Fig.\ \ref{fig-nets-examples}.  Meanwhile, the modules in the sparse structure can be recognised as the elementary cliques. Notably, the presence of a large clique increases the efficiency of building a nontrivial
structure, even for a small attractive potential, cf.\ Fig.\
\ref{fig-nets-examples} top. We will further discuss the community structure of these networks in connection with
their spectral properties in Sec.\ \ref{sec-spectra}. 
In Table\ \ref{tab-properties-nu} we show different graph measures of some mono-disperse assemblies whose
spectral properties are studied in  Sec.\ \ref{sec-spectra}. 
We note that the self-assembly process of cliques can result in a broad range of the degree of vertices. Depending on the size of
cliques $n\geq 3$, several hubs and a  power-law tail can appear at the sufficiently strong attraction between them \cite{we-SciRep2018}. For illustration, Fig.\ \ref{fig-ranking-p9} shows the ranking distribution of the degree for several monodisperse assemblies in the case of strong attraction.

\begin{figure}[!htb]
\begin{tabular}{cc} 
\resizebox{16pc}{!}{\includegraphics{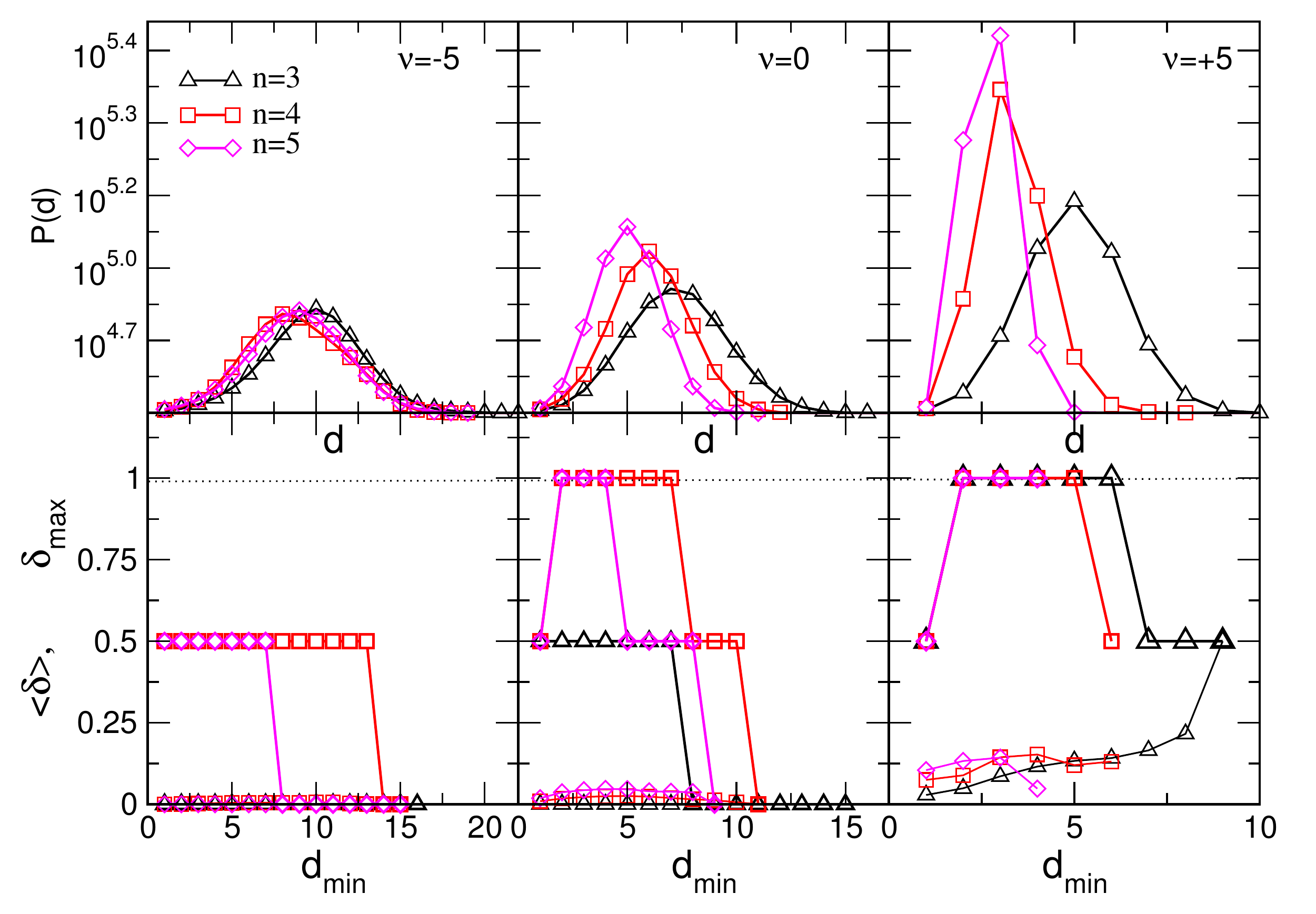}}\\
\end{tabular}
\caption{Shortest-path distance distributions $P(d)$ vs the distance $d$ and hyperbolicity
  parameter $\delta_{max}$ and $\langle \delta \rangle$ vs $d_{min}$
  for the assemblies of  simplexes indicated in the legend, and three distinct
  values of the chemical affinity parameter $\nu$.}
\label{fig-Hb-topol} 
\end{figure}

\begin{table}
\caption{Graph measures of the assemblies of cliques of the size $n$
  with $N\approx 1000$ vertices 
  for three representative values of the affinity parameter $\nu$:
  The number of egdes $E$, average degree $\langle k\rangle$ and clustering coefficient $\langle
  Cc\rangle$, graph's modularity $mod$, diameter $D$
  and the ratio of the hyperbolicity parameter
  $\delta_{max} $ to $D/2$. Two bottom rows are for mixed clique sizes
  $n\in[3,6]$ distributed according to $\varpropto n^{-\alpha}$.}
\begin{tabular}{|c|c|ccccccc|}
\hline
bb      &$\nu$&$E$& $<k>$&$<Cc>$& $<\ell >$&mod &D&$\delta_{max}/D/2$\\
\hline
          &-5&1501&2.999 &0.766&9.789 &0.928 &22&1/11\\
$n=3$&0&1734&3.465 &0.741 &7.265 &0.902 &16 &1/8 \\
          & +5&1991&3.982 & 0.735&4.958 &0.861 & 10& 1/5\\
\hline
         &-5&2009&4.61 &0.847&8.718  &  0.927&19&2/19\\
$n=4$&0&2426&4.852 & 0.808& 6.023& 0.895& 12& 1/6\\
          & +5&2984&5.968 &0.813 &3.23 &0.715 & 8& 1/4\\
\hline
          &-5&2514&  5.013 &0.878 &8.89  &0.921   &19&2/19\\
$n=5$& 0&3182&  6.351& 0.829 & 5.01 & 0.856   & 11& 2/11\\
         & +5&3997& 7.958& 0.850 &  2.703&0.850&     5& 2/5\\
\hline
\hline
$\alpha =2$&+5&2905&5.810&0.820&3.172&0.620&7&2/7\\
$\alpha =0$&+5&3464&6.298&0.844&2.857&0.569&6&1/3\\
\hline
\end{tabular}
\label{tab-properties-nu}
\end{table}

As mentioned above, the assemblies of cliques possess a negative curvature in the graph
metric space, which implies that they fulfil the  Gromov 4-point
hyperbolicity criterion \cite{HB-spaces-decomp2004}. More precisely, the graph $G$ is
hyperbolic \textit{iff} there is a constant $\delta (G)$ such that for any four vertices $(a,b,c,d)$,  the
relation $d(a,b)+d(c,d)\leq d(a,d)+d(b,c)\leq d(a,c)+d(b,d)$ implies that
\begin{equation}
\delta (a,b,c,d)=\frac{d(a,c)+d(b,d) -d(a,d)-d(b,c)}{2} \leq \delta (G) \ ,
\label{eq-HB}
\end{equation}
where $d(u,v)$ indicates the shortest path distance.  Note that the 
difference in  (\ref{eq-HB}) is bounded from above by the minimum
distance in the smalest sum $d_{min}\equiv
\mathrm{min}\{d(a,b),d(c,d)\}$. Thus, by plotting $\delta
(a,b,c,d)$ against $d_{min}$ for a large number of 4-tuples, we 
numerically estimate $\delta (G)\equiv \delta_{max}$ as the maximum
value observed in the
entire graph.

As described in Sec.\ \ref{sec-clnets}, the cliques aggregate by sharing their faces, i.e., cliques of a lower order, which leads to some specific properties of the grown structures \cite{we-SciRep2018}.  In particular,  the order of the simplicial complex cannot exceed the order of the largest attaching clique. Moreover, theoretical investigations of these types of
structures predict
\cite{HB-BermudoHBviasmallerGraph2013,HB-Bermudo2016,Hyperbolicity_chordality2017,Hyperbolicity_cliqueDecomposition2017}
that the upper bound of the hyperbolicity parameter of the graph differs from the
hyperbolicity of the "atoms" of the structure by at most one unit, that
is  $\delta_{max}=\delta_{a}+1$. Given that a clique is ideally
hyperbolic (i.e., tree-like in the shortest path metric space), we have
$\delta_{a}=0$, which gives $\delta_{max}=1$ for all clique complexes grown by the rules of our model. By sampling up to $10^9$ 4-tuples of vertices and computing the graph hyperbolicity parameter $\delta(G)$ in Eq.\ (\ref{eq-HB}), we demonstrate that the hyperbolicity parameter remains $\delta(G) \leq 1$ for all studied assemblies. More precisely, while
the structure of different assemblies, as well as their distribution of the shortest-path distances,  vary with the
chemical affinity $\nu$, the upper bound of their hyperbolicity
parameter remains fixed in agreement with the theoretical prediction.  In Fig.\ \ref{fig-Hb-topol}, we show the results of the numerical analysis for three representative sets of the assemblies of cliques of different sizes.  See also Table\ \ref{tab-properties-nu}.

\section{Spectral analysis of monodisperse
  assemblies\label{sec-spectra}}
\begin{figure}[!htb]
\begin{tabular}{cc} 
\resizebox{18pc}{!}{\includegraphics{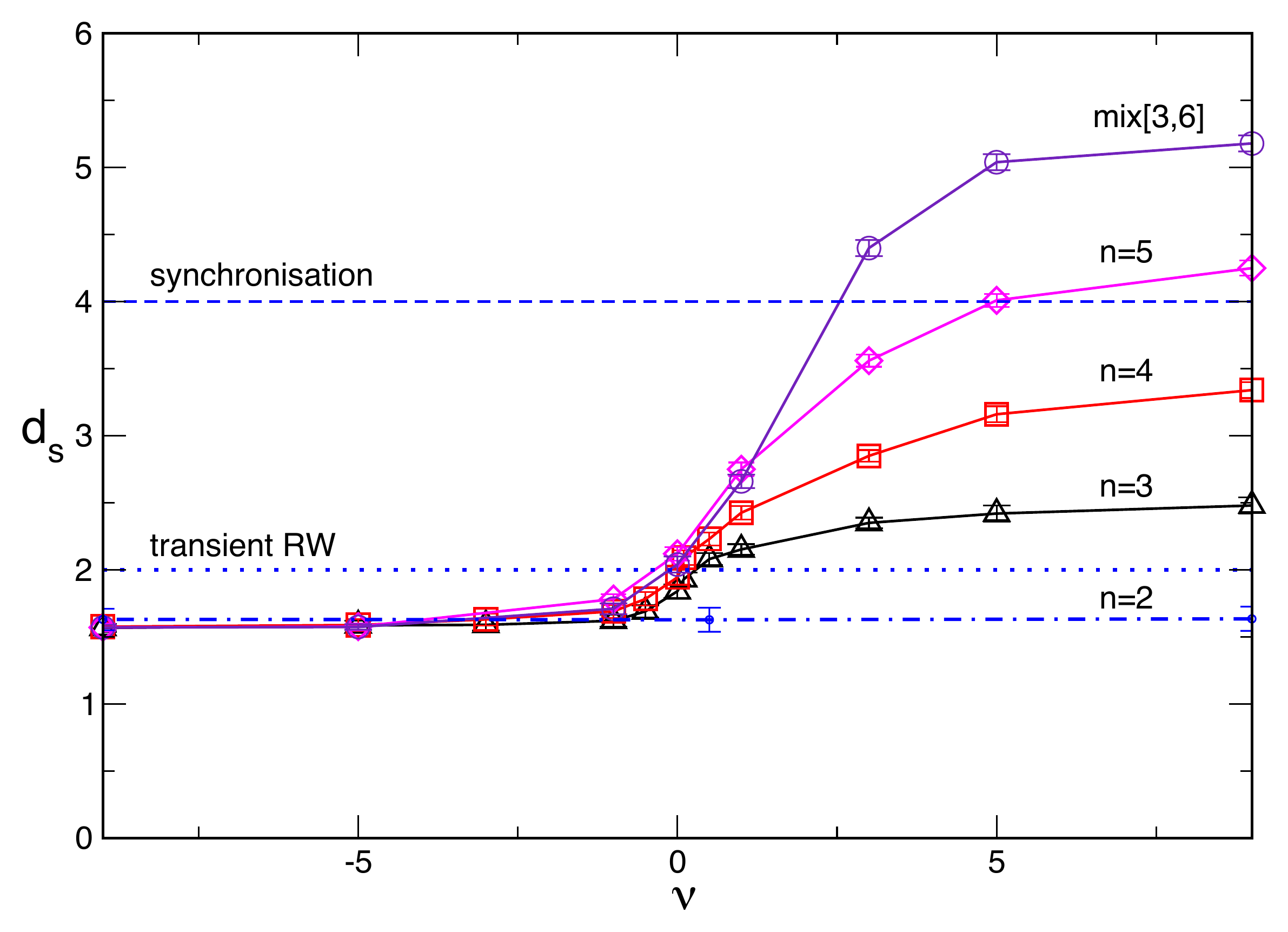}}\\
\end{tabular}
\caption{ The lines with different sysmbols represent the spectral dimension $d_s$ plotted against chemical affinity
  $\nu$ for the aggregates of monodisperse cliques of sizes $n=3,4,5$
  and a mixture of cliques of different sizes in the range
  $n\in[3,6]$. The bottom line corresponds to the random tree case $n=2$.
}
\label{fig-ds-nu} 
\end{figure}

\begin{figure*}[!htb]
\begin{tabular}{ccc} 
\resizebox{34pc}{!}{\includegraphics{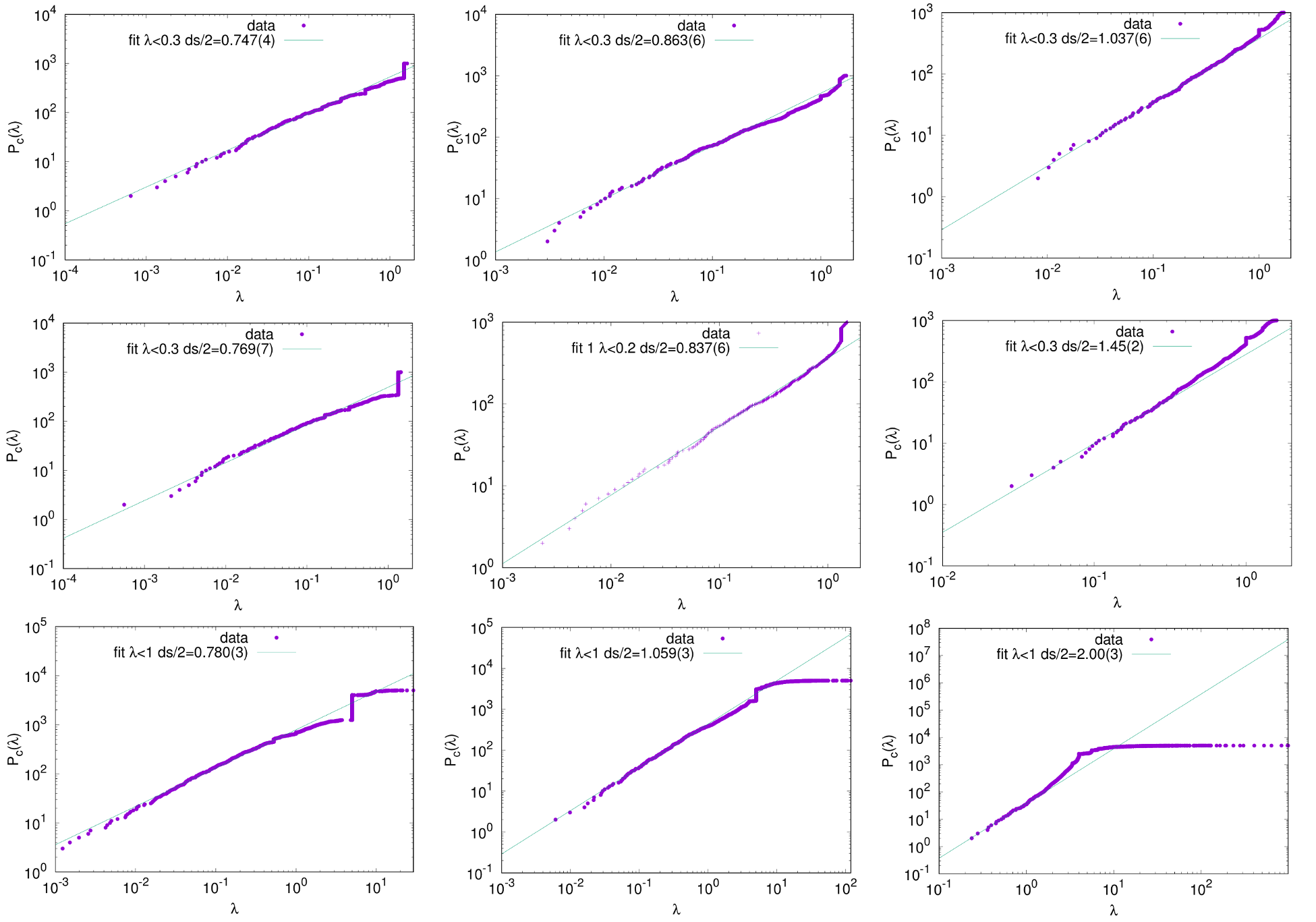}}\\
\end{tabular}
\caption{ Several examples of the cumulative spectral density
  $P_c(\lambda)$ in the range of small $\lambda$ for the
  Laplacian operator (\ref{eq-lp}) for the aggregates of triangles
  (top row), tetrahedra (middle row) and 5-cliques 
  (bottom row)  for varied chemical affinity $\nu=$ -5, 0, and +5,
  left-to-right column. }
\label{fig-fits-nu} 
\end{figure*}

\begin{figure*}[!htb]
\begin{tabular}{cc} 
\resizebox{28pc}{!}{\includegraphics{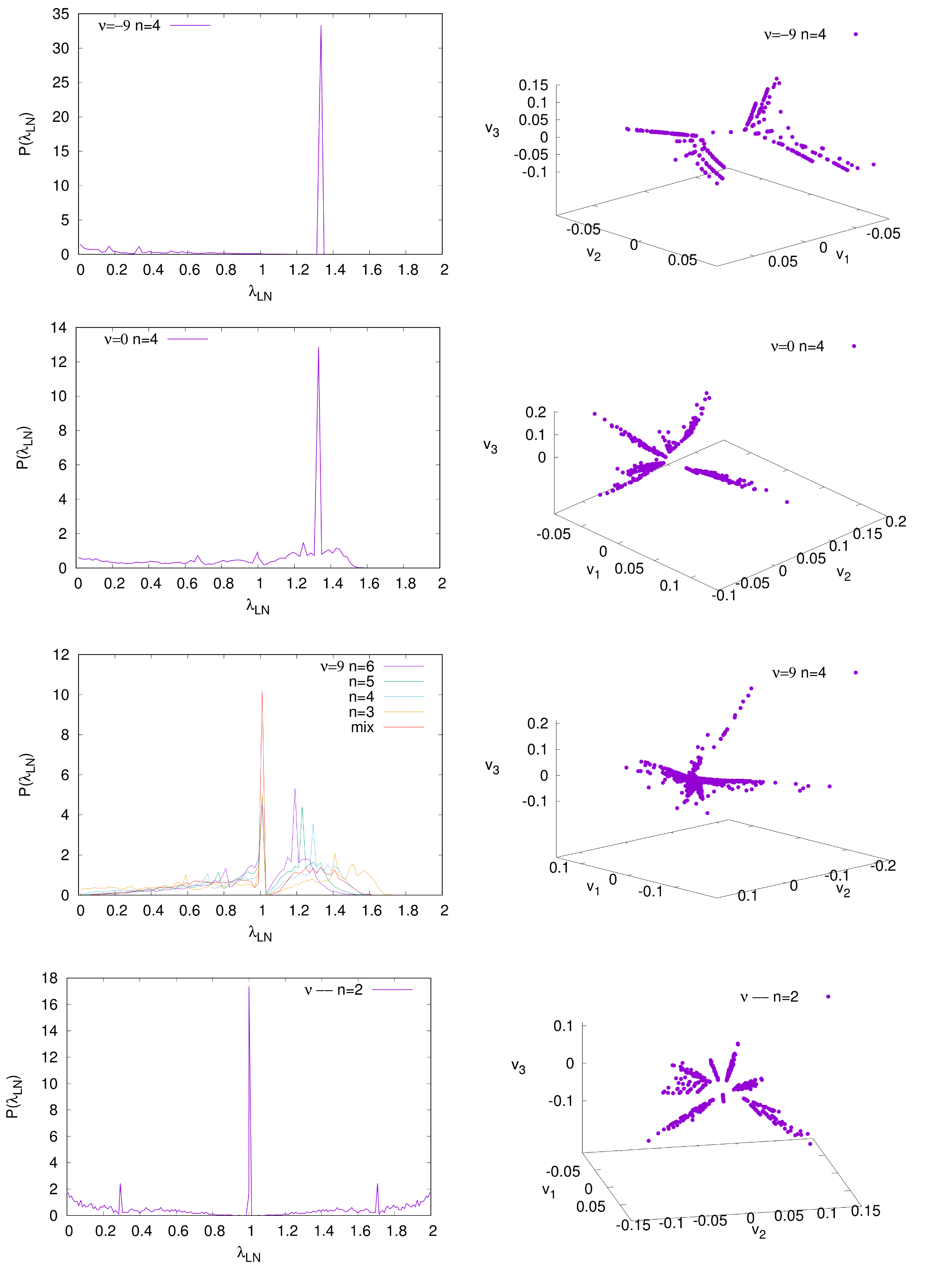}}\\
\end{tabular}
\caption{Spectral distribution (left column) and the corresponding scatter
  plots of
  the eigenvectors of the lowest nonzero eigenvalues of  the
  normalized Laplacian (right column)  for the
  aggregates of tetrahedra  (a-c,a1-c1) for $\nu$=-9, 0, and +9, and
  for the random tree structure $n=2$, which is independent of $\nu$, (d,d1). }
\label{fig-spectralDistr-nu} 
\end{figure*}

Spectral dimension $d_{s}$ of a graph, which is defined via
$\lim_{t\to\infty}\frac{\mathrm{log} P_{ii}(t)}{\mathrm{log}t}=-\frac{d_s}{2}$,  
characterises the distribution of return time $P_{ii}(t)$
of a random walk on that graph
\cite{rammal1983random,burioni1996universal,burioni2005random,
  durhuus2007spectral}. The diffusion type of processes
on network are described by Laplacian operators
\cite{Spectra-SFnetsSergey2003,Spectra-wePRE2009}. More precisely, for
the  undirected graph of $N$ vertices, two types of diffusion
operators are defined, i.e., the Laplacian operator with the components
\begin{equation}
L_{ij}=k_{i}\delta_{ij}-A_{ij} \ ,
\label{eq-lp}
\end{equation}
and the symmetric normalised Laplacian \cite{Spectra-Sergei2008} 
\begin{equation}
L^{n}_{ij}=\delta_{ij}-\frac{A_{ij}}{\sqrt{k_{i}k_{j}}} \ .
\label{eq-lpn}
\end{equation}
Here, $A_{ij}$ are the matrix elements of the adjacency matrix,
$k_{i}$ is the  degree of the node $i$, and
$\delta_{ij}$ is the Kroneker symbol. The operators defined with Eq.\ (\ref{eq-lp}) and (\ref{eq-lpn}) are
symmetric and have  real non-negative eigenvalues. Both operators
have the eigenvalue $\lambda=0$ with the degeneracy that is equal to the number of
connected components in the network. For the networks that have a finite
spectral dimension, spectral densities of both Laplacians scale as $P(\lambda)\simeq \lambda^{\frac{d_{s}}{2}-1}$
for small values of $\lambda$. Therefore, the corresponding cumulative distribution $P_{c}(\lambda)$ scales as
\begin{equation}
P_{c}(\lambda)\simeq \lambda^{\frac{d_{s}}{2}} \ ,  \quad  \lambda \ll 1 \,
\label{eq-spc}
\end{equation}
and it is suitable \cite{millan2019synchronization} for estimating the
spectral dimension $d_{s}$ of the network.
Here, we analyse the spectral properties of both  Laplacian operators
(\ref{eq-lp}) and (\ref{eq-lpn}) for the networks grown with different
building blocks and varied chemical affinity $\nu$, see Figures\ \ref{fig-ds-nu}-\ref{fig-spectralDistr-nu}.

We analyse the cumulative spectral density $P_{c}(\lambda)$ for the Laplacian defined
by the expression (\ref{eq-lp}) to determine the spectral dimension of
the graphs with the adjacency matrix $A_{ij}$. Note that the spectrum
is bounded from below, i.e., $0\leq\lambda$ for all eigenvalues
$\lambda$. According to Eq. (\ref{eq-spc}), we estimate $d_{s}$ for
each sample by fitting the data of $P_{c}(\lambda)$ for the values in the range
$\lambda\lesssim 0.3$, as illustrated  in  Fig.\ \ref{fig-fits-nu}.
The error bars are determined by taking the average from 
different samples of networks that have $1000$ and $5000$ nodes.
 The results summarised in Fig.\ (\ref{fig-ds-nu}) show how the spectral dimension of the corresponding
graphs varies with the chemical affinity $\nu$ depending on the size
of the elementary building blocks.

As Fig.\ \ref{fig-ds-nu} shows, the impact of the size of the
cliques strongly depends on the way that they
aggregate, which is controlled by the chemical
affinity  $\nu$. Precisely, for the sparse structures grown under the
considerable repulsion between 
cliques when $\nu < 0$, we find that the spectral dimension is  practically independent of
the size of cliques until the repulsion becomes vanishingly weak. In
contrast, when $\nu \geq 0$ the spectral dimension increases with the
size of the elementary cliques.  Here, the attaching cliques can share their larger
faces thus increasing the impact of the geometrical compatibility
factor. Remarkably, the spectral dimension increases with the strength
of the attraction between cliques, which favours sharing increasingly larger
faces. These faces are limited by the size of the elementary cliques. More
specifically, for all $\nu \geq 0$ values, $d_s$ is systematically
larger  in the aggregates of tetrahedra than those of triangles. In both
cases we have that $d_s$ exceeds the  limit of the transient random
walk, $d_s=2$,  for relatively  weak attraction between cliques $\nu \sim 1$.
However, both curves remain  below $d_s=4$ for the entire range of
$\nu$ values. Note that $d_s>4$ is
recognised as the full synchronisation condition for the Kuramoto oscillators
on network \cite{millan2019synchronization}. Whereas, in the region
$d_S\in (2,4]$ an entrained synchronisation with a complex
spatio-temporal patterns can be expected \cite{millan2019synchronization,Synchronization-frustratedBrain2014}.
 Even though a quite compact structure is grown by attaching tetrahedrons via their
triangular faces, see bottom panel in Fig.\ \ref{fig-nets-tetra2x}, its spectral
dimension remains limited  as $d_s< 4$, enabling the complex synchronisation patterns. 
We find that the limit $d_s=4$,  can be exceeded when the size of the clique is at
least $n=5$ and the attraction is considerably large, i.e., $\nu \geq
5$.  In this situation, the agglomerate consists of 5-cliques sharing many tetrahedrons
as their largest faces. Interestingly, it suffices to have a few
cliques of a large size to grow such agglomerates that cause the
spectral dimension  $d_s\geq 4$. For example, 
the  mixture shown in the top panel of Fig.\ \ref{fig-nets-examples} with  $n\in [3,6]$, where the population of 6-cliques is
only 1/4 of the population of 3-cliques, leads to the spectral
dimension shown by the top line  in Fig.\ \ref{fig-ds-nu}.
Furthermore, Fig.\
\ref{fig-fits-nu} indicates that not only the spectral dimension
but the entire spectrum changes with the size of the cliques and the
chemical affinity, as we discuss in more detail in the following.

Next, we determine the spectral density of the normalised Laplacian, defined by
(\ref{eq-lpn}), by averaging over $10$ networks of size $N\approx
1000$ generated for the same values of the model parameters. Note that
the eigenvalues of the normalized Laplacian are bounded in the range
\cite{Spectra-SFnetsSergey2003,Spectra-wePRE2009}   $\lambda_{LN}\in [0,2]$.  In Fig.\ \ref{fig-spectralDistr-nu}, we show the spectral
density of the normalised Laplacian for several representative cases, in particular, for  three different aggregates of tetrahedrons corresponding to the strong repulsion,
vanishing interaction, and strong attraction, respectively. Besides, in the panel (c), the spectral distribution is shown for the case of strong attraction $\nu=9$ for the cliques of different sizes $n\geq 3$.
 It should be noted that iso-spectral structures are observed in the
 case of the significant repulsion between the cliques $\nu=-9$. In
 this limit, apart from a structure at small eigenvalues, there is a
 prominent peak at $\lambda_{LN}=n/(n-1)$, indicating the presence of
 minimally connected cliques. In contrast, for $\nu\geq 0$, the attraction
between cliques and the relevance of
the geometrical compatibility factors lead to the appearance of larger simplicial complexes.
A peak starts building at $\lambda_{LN}=1$ already at $\nu=0$, which
gradually increases with the increasing $\nu$, as shown in
the panel (c). The occurrence of the peak at $\lambda_{LN}=1$,
(i.e., $\lambda =0$ in the corresponding adjacency matrix \cite{Spectra-Lap-Adj}) appears
as a characteristic feature of these hyperbolic networks. 
According to
previous studies of scale-free and modular networks
\cite{Spectra-SFnetsSergey2003,Spectra-wePRE2009}, this peak is
related to the nodes of the lowest degrees in the network. 

In the present study, such nodes are found in the bottom-right corner of the ranking distribution in Fig.\ \ref{fig-ranking-p9}. 
Apart from the random tree case, the appearance of this peak reflects the fact  that with the increased chemical
affinity a broad distribution of degrees occurs with a power-law tail, cf.\ Fig.\ \ref{fig-ranking-p9}.
 Notably, the highest peak is when the building cliques are of
different sizes $n\in [3,6]$, compared to the monodisperse structure
with $n=6$. A further exciting feature of these spectral densities is
that a characteristic minimum appears between $\lambda_{LN}=1$ and the
structure above it. The results in previous investigations
\cite{Spectra-hierarchPRE2005} suggest that such minimum in the
spectral density is a signature of the hierarchically organised
network. In the present study,  the hierarchical organisation of
cliques into simplicial complexes occurring at $\nu \geq 0$ has been
demonstrated by the algebraic topology methods in \cite{we-SciRep2018}. Here, we show by the spectral analysis that these simplicial complexes make the inner structure of mesoscopic communities, which can be identified by the localisation of the eigenvectors of the lowest non-zero eigenvalues \cite{Spectra-wePRE2009}.
In the right column of Fig.\ \ref{fig-spectralDistr-nu}, we show the scatter plot of the three eigenvectors related
to the lowest nonzero eigenvalues corresponding to the aggregates of tetrahedrons in the left column.  In the limit of strong repulsion between the cliques, the modularity of the entire structure is determined by the original cliques, see top right panel of Fig.\ \ref{fig-spectralDistr-nu}. Whereas,
the larger communities with sub-communities appear for $\nu \geq 0$ where higher-order connections  are increasingly more effective,  cf.\
panels (b1)  and (c1) of Fig.\ \ref{fig-spectralDistr-nu}. The bottom
panel corresponds to the random tree structure ($n=2$).

\section{Discussion and Conclusions\label{sec-discussion}}
We have studied topological and spectral properties of classes of
hyperbolic nano-networks grown by the cooperative self-assembly. The growth rules \cite{we-SciRep2018} that can be tunned by changing the parameter of chemical affinity $\nu$ enable to investigate the role of higher-order connectivity in the properties of the emerging structure. Attaching groups
of particles are parameterised by simplexes (cliques) of different
sizes which share a geometrical sub-structure by docking along with the growing network.
 For the negative values $\nu <0$, the repulsion among cliques makes them share a single node rather than an edge or a higher structure. On the other hand, $\nu \geq 0$ implies that the
geometrical factors and the size of the attaching clique become
relevant. In particular, the higher positive value of $\nu$  implies
that a new clique attaches to a previously added clique by sharing its
face of the larger order, thus building a more compact structure. 
Mathematically \cite{Hyperbolicity_cliqueDecomposition2017}, the attachment of cliques by sharing a face (of any order) leads to simplicial complexes whose hyperbolicity parameter cannot exceed one.  

Our results revealed that, while the hyperbolicity parameter remains fixed $\delta_{max}=1$ across different assemblies, their topological and spectral properties change with the increased chemical affinity,
see Table\ \ref{tab-properties-nu} and Fig.\ \ref{fig-ds-nu} and
Fig.\ \ref{fig-spectralDistr-nu}. Remarkably, the spectral dimension of the structure of strongly-repelled cliques of any size is practically indistinguishable from the one of a random tree of the same number of vertices. However, the rest of the spectrum is
different from the one of the tree structure; its dominant feature is the presence of cliques as the prominent network modules.
On the other hand, the compelling attraction between the
cliques for $\nu \gtrsim 0$ results in the spectral dimension that for
all sizes $n\geq 3$ exceeds the limit $d_s=2$, compatible with the
transient random walk on the network. 
Further increase of the spectral dimension with the increased affinity parameter $\nu$ strongly depends on the size of the cliques. Our results suggest that for a strong attraction with the cliques of size $n\geq 5$, the spectral dimension of the network can exceed the limit $d_s=4$, above which the synchronised phase is expected to exist \cite{millan2019synchronization}. However, more interesting structures are grown by smaller cliques or a mixture of different clique sizes with a weak attraction (small positive values of the parameter $\nu$) allowing the sharing a variety of clique's faces. In these cases, we find that the spectral dimension remains in the range of $d_s\in (2,4]$. These spectral properties are expected to be compatible with an entrained synchronisation \cite{millan2019synchronization} or a frustrated
hierarchical synchronisation with intricate spatiotemporal patterns
\cite{Synchronization-frustratedBrain2014}. A detailed analysis of
such synchronisation patterns on these graphs as well as potentially
super-diffusive processes \cite{we-superdiffusion}  remains for future work.
Due to their spectral properties, these structures can be interesting for modelling the complex dynamics in a variety
of biological systems and for potential applications. In the framework of the cooperative self-assembly of nanoparticle groups, our analysis shows how the control of the chemical affinity can lead to complex structures with different functional properties. 
Furthermore, the presented results can be relevant for a deeper
understanding of the functional complexity of many important structures
with built-in simplicial complexes, such as human connectome \cite{we-Brain-Connectome2019} and
other hierarchically modular networks.

\section*{\normalsize Acknowledgments}

The authors acknowledge the financial support from the Slovenian
Research Agency (research code funding number P1-0044)
and  from the Ministry of Education, Science and Technological Development of
the Republic of Serbia, the Projects ON171017 and by the Ito
Foundation fellowship.

\bibliographystyle{unsrt}

\end{document}